\documentstyle[12pt]{article}
\begin{document}

\def\a{\alpha}
\def\b{\beta}
\def\ch{\chi}
\def\d{\delta}
\def\e{\epsilon}
\def\f{\phi}
\def\g{\gamma}
\def\h{\eta}
\def\i{\iota}
\def\j{\psi}
\def\k{\kappa}
\def\l{\lambda}
\def\m{\mu}
\def\n{\nu}
\def\o{\omega}
\def\p{\pi}
\def\q{\theta}
\def\r{\rho}
\def\s{\sigma}
\def\t{\tau}
\def\u{\upsilon}
\def\x{\xi}
\def\z{\zeta}
\def\D{\Delta}
\def\F{\Phi}
\def\G{\Gamma}
\def\J{\Psi}
\def\L{\Lambda}
\def\O{\Omega}
\def\P{\Pi}
\def\S{\Sigma}
\def\U{\Upsilon}
\def\X{\Xi}
\def\T{\Theta}
\def\vf{\varphi}

\def\Ab{\bar{A}}
\def\gi{g^{-1}}
\def\li{{ 1 \over \l } }
\def\lb{\l^{*}}
\def\zb{\bar{z}}
\def\ub{u^{*}}
\def\vb{v^{*}}
\def\Tb{\bar{T}}
\def\pp {\partial }
\def\pb {\bar{\partial }}
\def\be{\begin{equation}}
\def\ee{\end{equation}}
\def\ben{\begin{eqnarray}}
\def\een{\end{eqnarray}}

\hsize=16.5truecm
\addtolength{\topmargin}{-0.8in}
\addtolength{\textheight}{1in}
\hoffset=-.5in

\thispagestyle{empty}
\begin{flushright} \ September \ 1995\\
SNUTP 95-100 \\
hep-th/9509120 \\
\end{flushright}
\begin{center}
 {\large\bf Complex sine-Gordon Theory \\
 for Coherent Optical Pulse Propagation  }\\[.1in]
\vglue .5in
 Q-Han Park\footnote{ E-mail address; qpark@nms.kyunghee.ac.kr }
\\[.2in]
{and}
\\[.2in]
H. J. Shin\footnote{ E-mail address; hjshin@nms.kyunghee.ac.kr }
\\[.2in]
{\it
Department of Physics \\
and \\
Research Institute of Basic Sciences \\
Kyunghee University\\
Seoul, 130-701, Korea}
\\[.2in]
{\bf ABSTRACT}\\[.2in]
\end{center}
It is shown that the McCall-Hahn theory of self-induced  transparency
in coherent optical pulse propagation can be identified with the
complex sine-Gordon theory in the sharp line limit. We reformulate the
theory in terms of the deformed gauged Wess-Zumino-Witten sigma model and
address
various new aspects of self-induced transparency.
 \vglue .1in

\newpage
Self-induced transparency(SIT), a phenomenon of anomalously low
energy loss in coherent optical pulse propagation, was first discovered
by McCall and Hahn\cite{McCall} and  the integrability of the SIT equation
 was demonstrated by employing the inverse scattering method\cite{AKN}.
When phase variation is ignored in the case for a symmetric frequency
distribution $g(\D w)$ of inhomogeneous broadening,  McCall and Hahn have
proved an area theorem for pulse propagation. In the sharp line limit where
the frequency distribution is sharply peaked at the carrier frequency $w_{0}$
such that  $g(\D w) = \d (w-w_{0})$, the SIT equation reduces to the well-known
sine-Gordon
equation and the $2\pi $ area pulse becomes a 1-soliton  of the sine-Gordon
theory. However, when phase variation is included,
the area theorem no longer holds and the structure of SIT in general has not
been well understood except for the construction of explicit solutions by
the inverse scattering method\cite{AKN}\cite{lamb}. In particular, despite
its integrability, the SIT theory in the sharp line limit has not been
identified with a known 1+1 dimensional integrable field theory, which made
 a systematic understanding of SIT in terms of a lagrangian
field theory impossible.

The purpose of this Letter is to show that the SIT theory with phase
variation can be
identified with the complex sine-Gordon theory in the sharp line
limit. The complex sine-Gordon theory, a generalization of the sine-Gordon
theory
with a phase degree of freedom, can be reformulated in terms of a
nonlinear sigma model which is known as the integrably deformed gauged
Wess-Zumino-Witten(WZW) model associated with the coset
SU(2)/U(1)\cite{park}\cite{shin1}. This allows us to address various new
aspects of SIT
in terms of characteristics of the complex sine-Gordon theory; e.g.
topological v.s. non-topological solitons, local gauge symmetry,
the U(1)-charge conservation, the chiral symmetry and the Krammers-Wannier
duality for dark v.s. bright solitons.
We also explain the off-resonance effect and inhomogeneous broadening of SIT
in the context of the local gauge symmetry of the present formulation.

The SIT equation is given by
\ben
\pb E + 2 \b <P> &=& 0 \nonumber \\
\pp D - E^{*}P - EP^{*} &=& 0 \nonumber \\
\pp P + 2i\D w P + 2ED &=& 0
\een
where $\D w = w-w_{0} \ , \ \pp \equiv \pp /\pp z \ , \ \pb \equiv \pp / \pp
\bar{z} \ ,
z= t-x/c , \bar{z} = x/c$.  $E,P$ and $ D$ represent the electric field,
the polarization and the population inversion respectively. The bracket
denotes an averaging over the distribution function of inhomogeneous
broadening. Since  Eq.(1) is invariant under the interchange
$(\b , E, P , D) \leftrightarrow (-\b ,E, -P , -D) $, we assume the coupling
constant $\b $ to be positive which we set to one by rescaling $E, P$ and
$ D$. In a simpler case where phase variation is ignored to make $ E$ real
and the frequency distribution is sharply peaked at
the carrier frequency($\D w =0$), we may parametrize $E, P$ and $ D$ by
\be
E=E^{*} = \pp \vf \ \ , \  \ <P> = P = -\sin{2\vf } \ \ , \ \
D= \cos{2\vf } \ .
\ee
Then the SIT equation reduces to the well-known sine-Gordon
equation
\be
\pb\pp \vf - 2\b \sin{2\vf } = 0 .
\ee
In order to include phase variation as well as the off-resonance
effect ($\D w \ne 0$), we assume $E$ to be complex and require that the
distribution is sharply peaked not necessarily at the carrier frequency, i.e.
$g(\D w) = \d (\D w - \x )$ for some constant $\x$.  Introduce a more general
parameterization
of $E, P$ and $D$ in terms of three scalar fields $\vf , \q $ and $\h $,
\be
E = e^{i(\q - 2\h )}( 2\pp \h {\cos{\vf } \over \sin{\vf }} - i\pp \vf )
\ \ , \ \
P = ie^{i(\q - 2\h )}\sin{2\vf }
\ \ , \ \
D = \cos{ 2\vf } \ .
\ee
The main result is that the SIT equation given in Eq.(1) then changes into a
couple of second order nonlinear differential equations known as the complex
sine-Gordon equation,
\ben
\pb\pp \vf + 4{\cos{\vf } \over \sin^{3}{\vf }}\pp \h \pb \h -
2\sin{2\vf } & =& 0
\\
\pb \pp \h - {2 \over \sin{2\vf }}(\pb \h \pp \vf + \pp \h \pb \vf
) &=& 0
\een
together with a couple of first order constraint equations,
\ben
2\cos^{2}{\vf }\pp \h - \sin^{2}{\vf }\pp \q - 2\x \sin^{2}{\vf }
&= & 0 \\
2\cos^{2}{\vf }\pb \h + \sin^{2}{\vf }\pb \q  &=& 0 \ .
\een

The complex sine-Gordon equation first appeared in 1976 in a description
of relativistic vortices in a superfluid \cite{lund}, and also
independently in a treatment of O(4) nonlinear sigma model\cite{pohl}.
Note that Eqs.(5) and (6) consistently reduce to the sine-Gordon equation
when phase variation is ignored so that $\h =0, \ \q = {\pi \over 2}$ and
the system is on resonance($\x = 0$). Earlier works on the complex
sine-Gordon theory have focused only on Eqs.(5) and (6). However, the
constraints in Eqs.(7) and (8) which are new expressions constitute an
essential part of the SIT theory, particularly in connection with a local
U(1)-gauge symmetry of SIT as explained later. Thus we will call
Eqs.(5)-(8) as the complex sine-Gordon theory. The gauge symmetry structure
as well as the integrability of Eqs.(5)-(8) may be best understood if we
reformulate the complex sine-Gordon theory in terms of the action principle.
The proper action is given by a group theoretical nonlinear sigma model
action known as the deformed gauged WZW action defined as follows;
\ben
S(g,A, \Ab , \b ) &=& S_{WZW}(g) +
{1 \over 2\pi }\int \mbox {Tr} (- A\pb g \gi + \Ab \gi \pp g
 + Ag\Ab \gi - A\Ab ) -S_{\mbox{potential}} \nonumber \\
S_{\mbox{potential}}&=& {\b \over 2\pi }\int \mbox{Tr}gT\gi \Tb
\een
where $S_{WZW}(g)$ is the conventional SU(2)-WZW action and $g $ is an
SU(2) matrix function. Tr denotes the trace and $ T = -\Tb = i\s_{3} =
\mbox{diag}(i, -i)$ for Pauli matrices $\s_{i} $. The local gauge fields
$A = a(z, \zb )\s_{3}, \Ab = \bar{a} (z,\zb )\s_{3} $  are introduced to
gauge the U(1) subgroup of SU(2). Owing to the absence of the kinetic terms,
the gauge fields $A, \Ab $ act as Lagrange multipliers which result in
the constraint equations. One of the nice properties of our fomulation is that
the equation of motion arising from the action (9) takes a zero
curvature form,
\be
\d_{g}S = -{1 \over 2\pi }\int \mbox{Tr}
[\  \pp + \gi \pp g + \gi A g + \b\l T \ , \ \pb + \Ab +
{1 \over \l }\gi \Tb g \ ]\gi \d g  = 0
\ee
where the constant $\l $ is a spectral parameter and the square bracket denotes
the
commutation. The constraint equations coming from the $A, \Ab $-variations are
\ben
\d _{A}S &=& {1 \over 2\pi }\int \mbox{Tr} ( \ - \pb g
\gi + g\Ab \gi - \Ab \  )\d A
= 0  \\
\d _{\Ab }S &=& {1 \over 2\pi }\int \mbox{Tr} ( \  \gi
\pp g  +\gi A g - A \ )\d\Ab = 0 \ .
\een
The action (9) is known to possess the local U(1)-vector gauge symmetry
under the transform;
\be
g \rightarrow h^{-1}gh \ \ , \ \ A \rightarrow A + h^{-1}\pp h \ \ , \ \
 \Ab \rightarrow \Ab  + h^{-1}\pb h
\ee
where $h = \exp (\phi (z, \zb) \s_{3})$, as well as the global U(1)-axial
vector
gauge symmetry under $g \rightarrow hgh $ for a constant $h$.
In order to identify Eqs.(10)-(12) with the SIT equation, we fix the vector
gauge by choosing
\be
A = \x T \ , \ \Ab =0
\ee
for a constant $\x $. Such a gauge fixing is possible due to the flatness of
$A, \Ab$\cite{shin1}. Also, we parameterize the $2\times 2$ matrix $g$ by
\be
g=e^{i\eta \s_{3}}e^{i\varphi (\cos{\q }\s_{1} -\sin{\q }\s_{2})}e^{i\eta
\s_{3}}= \pmatrix{ e^{2i\eta }\cos{\varphi } & i\sin{\varphi }e^{i\q } \cr
i\sin{\varphi }e^{-i\q } & e^{-2i\eta }\cos{\varphi } }\ .
\ee
Then the parametrization in Eq.(4) arises from an identification of
$E, P$ and $D$ with $g$ through the relation
\be
g^{-1}\pp g + \x g^{-1}Tg - \x T = \pmatrix{ 0 & -E  \cr E^{*} & 0 } \ \ , \ \
g^{-1}\bar{T}g = -i \pmatrix{ D & P \cr P^{*} & -D }
\ee
where we have used the constraint equation (12). Also, the zero
curvature equation (10) with the identification in Eq.(16) becomes
\be
\left[ \  \pp + \pmatrix{ i\b \l + i\x & -E \cr E^{*} & -i\b \l -i\x } \ , \
\pb - {i \over \l } \pmatrix{D & P \cr P^{*} & -D }
\right]  = 0
\ee
whose components agree precisely with the SIT equation in the sharp line
limit. The constraint
equations (11) and (12), combined with Eq.(14), also reduce to Eqs.(7)
and (8). Thus we have shown that the SIT equation consistently arises from the
action (9) with the gauge fixing in Eq.(14). Moreover, the zero curvature
equation (17) demonstrates the integrability of the SIT equation.
The potential term in Eq.(9) changes into the population inversion $D$,
\be
S_{\mbox{potential}} = \int {\b \over \pi }\cos{2\varphi } =
\int {\b \over \pi }D   ,
\ee
which for $\b =1 $ possesses degenerate vacuua at
\be
\varphi = \vf_{n} = (n+ {1\over 2} )\pi , \ n \in Z  \ \mbox{ and } \
\q \ = \q_{0} \ \ \mbox{for} \ \ \q_{0} \  \mbox{constant} \ .
\ee
The soliton solutions interpolating different vacuua can be obtained
either by applying the dressing method or by using the B\"{a}cklund
transformation\cite{shin1}.
In particular, the 1-soliton solution is given by
\ben
\cos{\vf } &=& {b \over \sqrt{(a-\x )^{2} + b^{2} }}
\mbox{sech} (2bz -2b C \zb ) \nonumber \\
\h &=& (a-\x )z + (a-\x ) C\zb \nonumber \\
\q &=& - \tan^{-1}[ {a-\x \over b } \mbox{coth} (2bz - 2bC\zb )] - 2\x z +
2D\zb
\een
where $a, b$ are arbitrary constants and
\be
C = {1 \over (a-\x )^{2} + b^{2}} \ , \ D = 0 .
\ee
In terms of $E$,
\be
E = -2ib \  \mbox{sech} (2bz - 2bC \zb ) e^{-2i(az -D\zb  + (a-\x )C\zb )} .
\ee
If $a-\x =0$, this solution interpolates between two different vacuua
$\vf_{n}$ and $\vf_{n+1}$, i.e. it becomes a topological 1-soliton
($\D n =1$).  On resonance where $a = \x =0$, Eq.(20) reduces to
the 1-soliton of the sine-Gordon equation, or a $2\pi $ pulse of SIT.
If $a -\x \ne 0$, the solution in Eq.(20) reaches to the same vacuum
asymptotically as $x \rightarrow \pm \infty $
so that the topological number is zero ($\D  n=0$).
Nevertheless, except for the topological number, this solution possesses
all the properties of a soliton so that we call it a nontopological
1-soliton. It represents a localized pulse with a steadily varying
phase. The time area of the pulse, which is defined by
\be
\D S = 2\int |E| dt ,
\ee
is still $2\pi $. It is important to mention that this $2\pi $ area is
a mere coincidence and should not be confused with the $2\pi $ area of
the topological one. Because of the interference between phases of each
nontopological solitons, multi-nontopological solitons in general
do not possess the area which is an integer multiple of $2\pi $
and the area theorem of McCall and Hahn in the case of inhomogenous
broadening does not hold.
The stability of nontopological solitons, unlike the topological case whose
stability is due to the topological protection, arises from the U(1)-charge
conservation law. Recall that the action (9), consequently Eqs.(5)-(8),
are invariant under the axial vector transform $g \rightarrow hgh$ or,
equivalently,
\be
\h \rightarrow \h + \e \ \   \mbox{ for } \e  \  \mbox{constant} .
\ee
The corresponding Noether currents and the charge are given by
\ben
J &=& {\cos^{2}{\vf } \over \sin^{2}{\vf }}\pp \h  \ \ , \ \
\bar{J} = {\cos^{2}{\vf } \over \sin^{2}{\vf }}\pb \h \nonumber \\
Q &=& \int_{-\infty }^{\infty }dx (J + \bar{J})
\een
where $J, \bar{J}$ satisfy the conservation law,
 $  \pp \bar{J} + \pb J = 0$.
In particular, the charge of the 1-soliton in Eq.(20) is
\be
Q_{\mbox{1-sol}} = -c(\mbox{sign}[b\cdot (a-\x )]{\pi \over 2 } -
\tan^{-1}{a-\x \over b}) .
\ee
The stability of nontopological solitons can be shown either by using
conservation laws in terms of charge and energy as given in \cite{shin1},
or by studying the behavior against small fluctuations\cite{shin3}.

The action (9) also possesses two different types of discrete
symmetries. This can be seen most easily in a different gauge where
$A=\Ab =0$ which is connected to the gauge in Eq.(14) by an appropriate
vector gauge transformation.
The first case is {\it the chiral symmetry} under the interchange,
\be
z \leftrightarrow \zb \  \ \mbox{and} \ \
 \ g \leftrightarrow g^{-1} ( \ \mbox{ or } \h \leftrightarrow -\h
\ , \ \vf \leftrightarrow -\vf )
\ee
which is a characteristic of the WZW action.
Unlike the case of a sigma model without the Wess-Zumino term,
parity alone $(z \leftrightarrow \zb $) is not a symmetry.
In the SIT context, this is due to the slowly varying envelop approximation
which breaks the
parity invariance of the Maxwell-Bloch equation. The chiral symmetry generates
a new solution from a known one. For example, the chiral transform
of the 1-soliton in Eq.(20) in the resonant case $(\x = 0)$
is again a 1-soliton but with the replacement of constants $a, b$ by
\be
a \rightarrow -{a \over a^{2} + b^{2}} \ , \
b \rightarrow { b \over a^{2} + b^{2}} ,
\ee
which changes the shape of the pulse as well as the velocity by
 $v \rightarrow c - v$.
The currents and the charge also change into
\be
J \rightarrow -\bar{J} \ , \ \bar{J} \rightarrow - J \ , \ Q \rightarrow -Q .
\ee
The second case is {\it the dual symmetry} of the
Krammers-Wannier type\cite{shin1} under
\be
\b \leftrightarrow -\b \ , \
g \leftrightarrow  i\s_{1}g \ .
\ee
Changing the sign of $\b $ makes the potential upside down so that
vacuua become maxima of the potential and vice versa. This allows us to find a
localized
solution which approaches to a maximum of the potential asymptotically,
i.e. it reaches to the completely population inverted state.
In analogy with the nonlinear Schr\"{o}dinger case, we call it a ``dark
soliton". In practice, the dark soliton for positive $\b $ can be obtained by
replacing $\b \rightarrow -\b \ , \ \zb \rightarrow -\zb $ in the (bright)
soliton
solution of the negative $\b $ case. For example, the dark 1-soliton can
be written by
\ben
\cos{\vf }e^{2i\h } &=& - {b \over \sqrt{(a-\x)^{2} + b^{2}}}\mbox{tanh}
(2bz + 2bC\zb ) - i{a-\x \over \sqrt{(a-\x )^{2} + b^{2}}} \nonumber \\
\q &=& -2(a- \x) (z- C\zb ) -2\x z
\een

Finally, we show that inhomogeneous broadening can be incorporated
into our formulation naturally with minor modifications.
We maintain the constraint equation (12) only and modify the zero curvature
eqution by
\be
\left[ \pp + g^{-1}\pp g + \x g^{-1}Tg - \x T + \tilde{\l }T \ , \ \pb +
\left< {g^{-1}\bar{T} g \over \tilde{\l } - \x } \right> \right] = 0
\ee
where the constant $\tilde{\l }$ is a spectral parameter which becomes $\l + \x
$
in the sharp line limit.  We make the same identification as in Eq.(16) and
require that $g^{-1}\pp g + \x g^{-1}Tg - \x T$ is  $\x $-independent since
$E$ is a $\x $-independent macroscopic quantity. This results in
the SIT equation with inhomogenous broadening as given in Eq.(1).
Once again,  by using the dressing method, the 1-soliton can be obtained
which is the same as in Eq.(20) but with the replacement
\be
C = \left< { 1\over (a-\x )^{2} + b^{2}} \right> \ ,
 \ D = (a-\x )\left< { 1\over (a-\x )^{2} + b^{2}} \right>
- \left< {  a-\x \over (a-\x )^{2} + b^{2}} \right> .
\ee
Eq.(14) shows that each frequency $\x $ corresponds to a specific gauge choice
in our formulation therefore inhomogenous broadening is equivalent to
averaging over different gauge fixings.
This implies that the inhomogenously broadened case can not be treated by a
single field theory. Nevertheless it is remarkable that the group
theoretical parametrization of $E, P$ and $ D$ is still valid.
Another important feature of inhomogenous broadening is that it introduces
an anomaly term $M$ in the U(1)-current conservation such that
$\pp \bar{J} + \pb J = M$ and
\ben
M &=& 2\cot{\vf } [ \ \cos (\q - 2\h ) < \sin (\q - 2\h ) \sin {2\vf } > -
\sin (\q - 2\h ) < \cos (\q - 2\h ) \sin {2\vf } >  \nonumber \\
&& -(\cot^{2}{\vf }\pb \h  + {1\over 2}\pb \q )\pp \vf   \ ] .
\een
This anomaly vanishes in the sharp line limit due to the constraint Eq.(8).
It also vanishes in the case of 1-soliton and the charge remains conserved.
This may be compared with the conserved area of topological solitons in the
presence of inhomogenous broadening. It is an open question whether there
exists a similar theorem to the area theorem concerning about the stability
of pulses with phase variation in terms of charge and anomaly.

We may choose different groups and coset structures for the deformed gauged
WZW action\cite{shin2} which in our group theoretical formulation of SIT
leads to other cases than the two level SIT.
These cases, together with other aspects of SIT which were not considered
in this Letter, will appear in a longer version of this Letter\cite{shin3}.
\vglue .3in
{\bf ACKNOWLEDGEMENT}
\vglue .2in
This work was supported in part by the program of Basic Science Research,
Ministry of Education BSRI-95-2442, and by Korea Science and Engineering
Foundation through CTP/SNU and Korea-Japan collaboration program.
\vglue .2in

\end{document}